\documentclass[preprint]{acmart}

\usepackage{booktabs} % For formal tables
\usepackage{amsmath}
\usepackage{amssymb}
\usepackage{amsopn}
\usepackage{tikz}
\usetikzlibrary{calc,trees,positioning,arrows,chains,shapes.geometric,%
    decorations.pathreplacing,decorations.pathmorphing,shapes,%
    matrix,shapes.symbols}

\tikzset{
>=stealth',
  punktchain/.style={
    rectangle,
    rounded corners,
    draw=black, very thick,
    text width=7em,
    minimum height=3em,
    text centered,
    on chain},
     connector/.style = {->,thick},
  line/.style={draw, thick, <-},
  element/.style={
    tape,
    top color=white,
    bottom color=blue!50!black!60!,
    minimum width=6em,
    draw=blue!40!black!90, very thick,
    text width=7em,
    minimum height=2.5em,
    text centered,
    on chain},
  every join/.style={->, thick,shorten >=1pt},
  decoration={brace},
  tuborg/.style={decorate},
  tubnode/.style={midway, right=2pt},
}

\acmConference[The Web Conference 2018]{The Web Conference}{April 2018}{Lyon, France}
\acmYear{2018}
\copyrightyear{2018}

\setcopyright{none}
\renewcommand\footnotetextcopyrightpermission[1]{} % removes footnote with conference information in first column
\begin{document}

\title{Etymo: A New Discovery Engine for AI Research}

\author{Weijian Zhang}
\affiliation{%
  \institution{The University of Manchester}
  \streetaddress{Oxford Rd}
  \city{Manchester}
  \postcode{M13 9PL}
}
\email{weijian.zhang@manchester.ac.uk}

\author{Jonathan Deakin}
\affiliation{%
  \institution{The University of Manchester}
  \streetaddress{Oxford Rd}
  \city{Manchester}
  \postcode{M13 9PL}
}
\email{jonathan.deakin@postgrad.manchester.ac.uk}

\author{Nicholas J. Higham}
\affiliation{%
  \institution{The University of Manchester}
  \streetaddress{Oxford Rd}
  \city{Manchester}
  }
\email{nick.higham@manchester.ac.uk}

\author{Shuaiqiang Wang}
\affiliation{%
  \institution{JD.com}
  \city{Beijing}
}
\email{wangshuaiqiang1@jd.com}

\def\R{\mathbb{R}}

\newcommand{\cosim}{\operatorname{cosine\_similarity}}
\newcommand{\tfidf}{\operatorname{tf\_idf}}
\newcommand{\df}{\operatorname{df}}
\newcommand{\tf}{\operatorname{tf}}
\newcommand{\idf}{\operatorname{idf}}

\def\norm#1{\|#1\|}

\begin{abstract}

  We present Etymo (\url{https://etymo.io}), a discovery engine to facilitate
  artificial intelligence (AI) research and development.
  It aims to help readers navigate a large number of
  AI-related papers published every week by using a novel form of search that
  finds relevant papers and displays related papers in a graphical interface.
  Etymo constructs and
  maintains an adaptive similarity-based network of research papers as an
  all-purpose knowledge graph for ranking, recommendation, and
  visualisation.
  The network is constantly evolving and can learn from user feedback to adjust itself.

\end{abstract}


\maketitle

%%%%%%%%%%%%%%%%%%%%%%
\section{Introduction}\label{sec:introduction-1}

The rapid growth of global scientific output
creates new challenges for information retrieval.  The problem is
particularly acute in AI (artificial intelligence) research. ArXiv
(\url{https://arxiv.og}), for example, gains around 500 new AI-related papers every week
and the number is growing. As a result, it is difficult for researchers to
keep up-to-date with the latest developments in AI research.  We have built
a new discovery engine for scholarly research called Etymo that addresses this challenge.

Citations of scientific papers are generally considered an important
indicator of a paper's impact \cite{hirsch05}.
Google Scholar's ranking algorithm is not publicly known, but
research \cite{beel09} has shown that citation counts have the highest
weighting in its ranking algorithm.  However, recent publications have
few or no citations so it is difficult to use citations to judge the
importance of very recent papers; thus, recent insightful publications are difficult to rank.
Our idea is to build a
similarity-based network and use
this information for information retrieval tasks.

How can we obtain links in a non-hypertext setting?  Our approach is to
infer links from the distributed vector representation of the full-text
papers, i.e., if the cosine similarity between the vector representations
is large, we link these two papers. It turns out that inferred links are
similar to the citation network because papers talking about the same
subject tend to cite one another.  However,
the analogy between hyperlinks
and generated links is not perfect. In particular, auto-generated links are
a noisier source of information and much more prone to spam.

He et al. \cite{he17} proposed a meta-approach called HICODE (HIdden COmmunity
DEtection) for discovering the hidden community structure in a network. By
removing certain edges from the network (weakening the strength of certain
structures), one can uncover other structures in the network.  Similarly,
we can strengthen and weaken the connectivity of the network structure
so as to improve our ranking algorithm and
filter out unwanted papers. We do this by exploiting papers' social media
activities (such as the number of retweets on Twitter)
and certain information from user feedback.
The resulting graph is used for ranking, recommendation, and visualisation.
Note also that inferred links can be generated faster than citations
(papers may take ~1 year to be cited, but inferred links can be generated almost instantly).

We use a combination of PageRank and Reverse PageRank for ranking
papers, which we find gives better search results than pure PageRank or
HITS \cite{kleinberg99}.  In Reverse PageRank, we compute PageRank on the
graph with reversed direction, i.e., reverse the direction of each edge
$(i,j)$ to $(j, i)$. Fogaras \cite{fogaras03} shows that Reversed Page Rank
scores express hub quality.
We have also designed and implemented a new search interface where we display
search results as a combination of a list and relationship
visualisation. This new interface allows readers to quickly locate relevant and related papers.

Our Etymo discovery engine provides a way to evaluate new
research papers by exploiting the full-text of research papers. It
challenges the traditional list-based search interface by combining an item
list with item relationship visualisation.
Etymo updates its database on a
daily basis and is free to use for all (demo available at \url{https://etymo.io}). Registered users can also write
notes and receive recommendations.

\section{Related Works}

Recently, researchers have realised that the full text of scientific papers
is an important resource for search and other applications.  Indeed,
Salatino et al. \cite{salatino16} use the semantic enhanced
topic network (where nodes are topics and edges are their co-occurrences in
a sample of publications) to identify the appearance of new topics.
Sateli
et al.\ \cite{sateli17} analyze full-text research articles to generate
semantic user profiles.
Semantic Scholar
(\url{https://www.semanticscholar.org})
\footnote{https://techcrunch.com/2016/11/11/scientists-gain-a-versatile-modern-search-engine-with-the-ai-powered-semantic-scholar/}
is a search engine for scholarly research that analyzes and links key
information from the full text of research papers for improving search
results.
Similarly, Etymo makes use of the full text of papers to generate a
similarity-based network, which is then used for information retrieval
tasks.

\section{Architecture Overview}

\begin{figure}
%\centering
\begin{tikzpicture}
  [node distance=.6cm,
  start chain=going below,]
     \node[punktchain, join, fill=white!30!red!50] (papers) {Research Papers};
     \begin{scope}[start branch=venstre,
       %We need to redefine the join-style to have the -> turn out right
       every join/.style={->, thick, shorten <=1pt}, ]
       \node[punktchain, on chain=going left, fill=white!30!red!50]
           (social) {Social Media};
     \end{scope}
     \node[punktchain, join] (crawler) {Crawler};
     % \node[punktchain, join] (store) {Store Server};
     \node[punktchain, join] (analysis) {Analysis};
     \begin{scope}[start branch=venstre,
       %We need to redefine the join-style to have the -> turn out right
       every join/.style={->, thick, shorten <=1pt}, ]
       \node[punktchain, on chain=going left]
           (lexicon) {Lexicon};
     \end{scope}
     \node[punktchain, join] (network) {Similarity-based Network};
      \node (ranking) [punktchain, join]  {Paper Ranking};
      \begin{scope}[start branch=venstre,
        %We need to redefine the join-style to have the -> turn out right
        every join/.style={->, thick, shorten <=1pt}, ]
        \node[punktchain, on chain=going left] (location) {Paper Location};
      \end{scope}
      \node(feed) [punktchain, join, fill=black!10!green!50]  {Feed};
      \begin{scope}[start branch=venstre,
        %We need to redefine the join-style to have the -> turn out right
        every join/.style={->, thick, shorten <=1pt}, ]
        \node[punktchain, on chain=going left, fill=black!10!green!50] (search) {Search};
      \end{scope}

  \path[connector] (lexicon) [bend right=55] edge  (search);
  \path[connector] (location) edge (search);
  \path[connector] (ranking) edge (feed);
    \path[connector] (ranking) edge (search);
  \path[connector] (network) edge (location);
  \path[connector] (location) edge (feed);
  \path[connector] (analysis) edge (lexicon);
  \path[connector] (analysis) edge (network);
 \path[connector] (network) edge (search);
 \path[connector] (network) [bend left=62] edge (feed);
 \path[connector] (network) edge (ranking);
 \path[connector] (social) edge (crawler);
  \end{tikzpicture}
  \caption{The dependency graph of Etymo. The direction of an arrow
    indicates the dependence of the two components connected by it. The red
    rectangles represent our data source; the green rectangles are the two
    main components of Etymo.} \label{fig:workflow}
\end{figure}
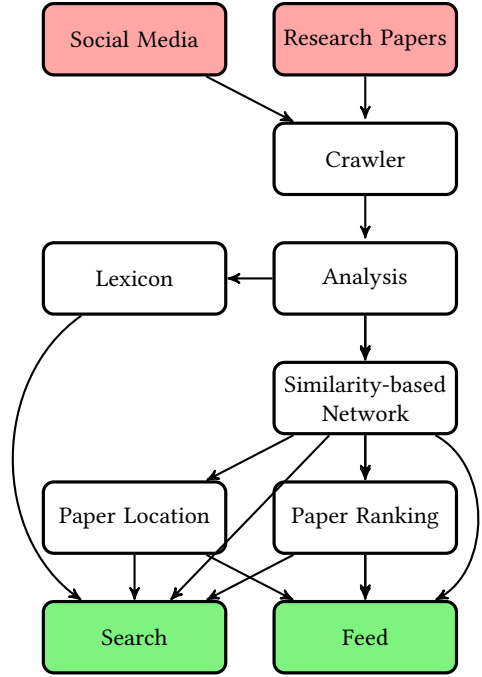

The dependency graph of Etymo is shown in Figure \ref{fig:workflow}.  Etymo
has several crawlers for downloading research papers from different journal
websites.
For each paper in our database, we store both the PDF version of the paper and the metadata, including author name, journal name, paper abstract, and the date of publication.

In the Analysis stage, we convert all the PDFs to text using pdftotext
\footnote{\url{https://en.wikipedia.org/wiki/Pdftotext}}. We then apply
Doc2Vec \cite{le14} and TF-IDF \cite[Chap 6]{manning08} to represent a document $d$ as numeric vectors
$v_{Doc2Vec}(d)$ and $v_{tf-idf}(d)$ respectively. Both algorithms
represent a paper by a numeric vector such that papers with similar content
are close to each other in the vector space.
This content similarity information is then
used for building a similarity-based network of all the papers in the
database.
We generate two networks using the two alorithms and find in practice a combination of the two networks can product better results than just using one of the two.
We use t-SNE \cite{maaten08} to find the paper locations and network centrality
algorithms for the paper ranking.  We also generate a lexicon from the
TF-IDF's global term weights, which is later used in search.
The main components of Etymo consist of a search engine and a feed
engine.
Results from both engines are displayed as a list and graph visualisation.

% \vskip0.2in

\section{System Features}

Etymo has two important features that help it produce useful search
results. First, it uses a document vector representation of the full-text
papers to build a similarity-based network, where papers are nodes and
similar papers are linked.  This network is adaptive because of a user
feedback mechanism: users' stars, clicks and Twitter mentions are used to
reinforce the `correct' connections and weaken the `unimportant' ones. The
resulting network is then used for ranking and recommendation. Second, we
have designed and implemented a novel search interface, with results
presented both in a traditional item list and with a visualisation showing paper
relationships in order to help users quickly find related papers and have
a general idea of a research area.

\subsection{Similarity-based Network}\label{sec:simil-based-netw}

To construct the similarity-based network, one first needs to represent
documents using numeric vectors. We use a distributed representation of the
documents called Doc2Vec \cite{le14} and a bag of words model called TF-IDF.
We then construct the similarity-based network using the cosine similarity
measure.  One potential problem with this similarity-based network approach
is that it does not distinguish a high quality paper from a bad one. We
argue that we can use user feedback to adjust the network structure in
order to give important papers higher weights.

\subsubsection{Adaptive Network and Ranking}\label{sec:build-simil-based}

Each paper is a node in the network and similar papers are linked together.
We compute the cosine similarity of every pair of paper vectors in the database.
If the cosine similarity score of two
papers' vector representations is larger than a given threshold, we link
these two papers.
In other words, if $u\in \R^n$ and $v \in \R^n$ (where $n = 1000$ in practice)
are the vector representation of two papers $p_u$ and $p_v$, we define
$$
\cosim(u, v) = \frac{\sum_{i=1}^n u_i v_i}{\norm{u} \norm{v}},
$$
where $\norm{u} = \left(\sum_iu_i^2 \right)^{1/2}$.  If
$\cosim(u, v) = w > \alpha$, where $\alpha$ is a
threshold, we link paper $p_u$ and $p_v$ by an edge with weight $w$.

Calculating the similarity scores of two
papers has a time complexity $O(n)$.
Adding a new paper when there are already $m$ papers in the network
therefore costs $O(mn)$, which is clearly prohibitive for large $m$.
One potential solution is to calculate a new paper's similarity on a representative
subset of the existing papers, i.e., find top $k$ high quality papers,
where $k \ll n$. Since our graph centrality ranking provides a measure of paper quality, we use the top $k$ high ranking papers as a representative subset.
Then for a new paper, we only calculate its similarity with these $k$ papers.

Similarity-based networks are vulnerable to spam. For example, if a paper
contains a large number of important key words in AI research, it may have
a high connectivity on the network hence a high score in the ranking.
We use authors' votes to adjust the structure of
the network in three main ways:
\begin{enumerate}
\item We use user stars to increase the edge weights to a
node, this increases the number of edges to that
node.
In other words, a paper with many user stars connects to more papers than a
paper with few.
\item We use user libraries to infer connectivities: increasing edge weights between the papers in a user's library.
\item We weaken the connectivities of a highly ranking paper if it has poor
click rates.
\end{enumerate}

Finally, we turn the undirected similarity-based network into a directed
network using the temporal information from the paper published date.
As a result, a new paper on this network `recommends' a old paper if they
have similar content or user data suggests that they are related.
Intuitively we attempt to predict the citation network structure of new research
papers when their citations are not available.

\subsection{Go Beyond List: Relationship Visualisation}\label{sec:go-beyond-list}

\begin{figure}[h]
\centering
\includegraphics[width=\textwidth]{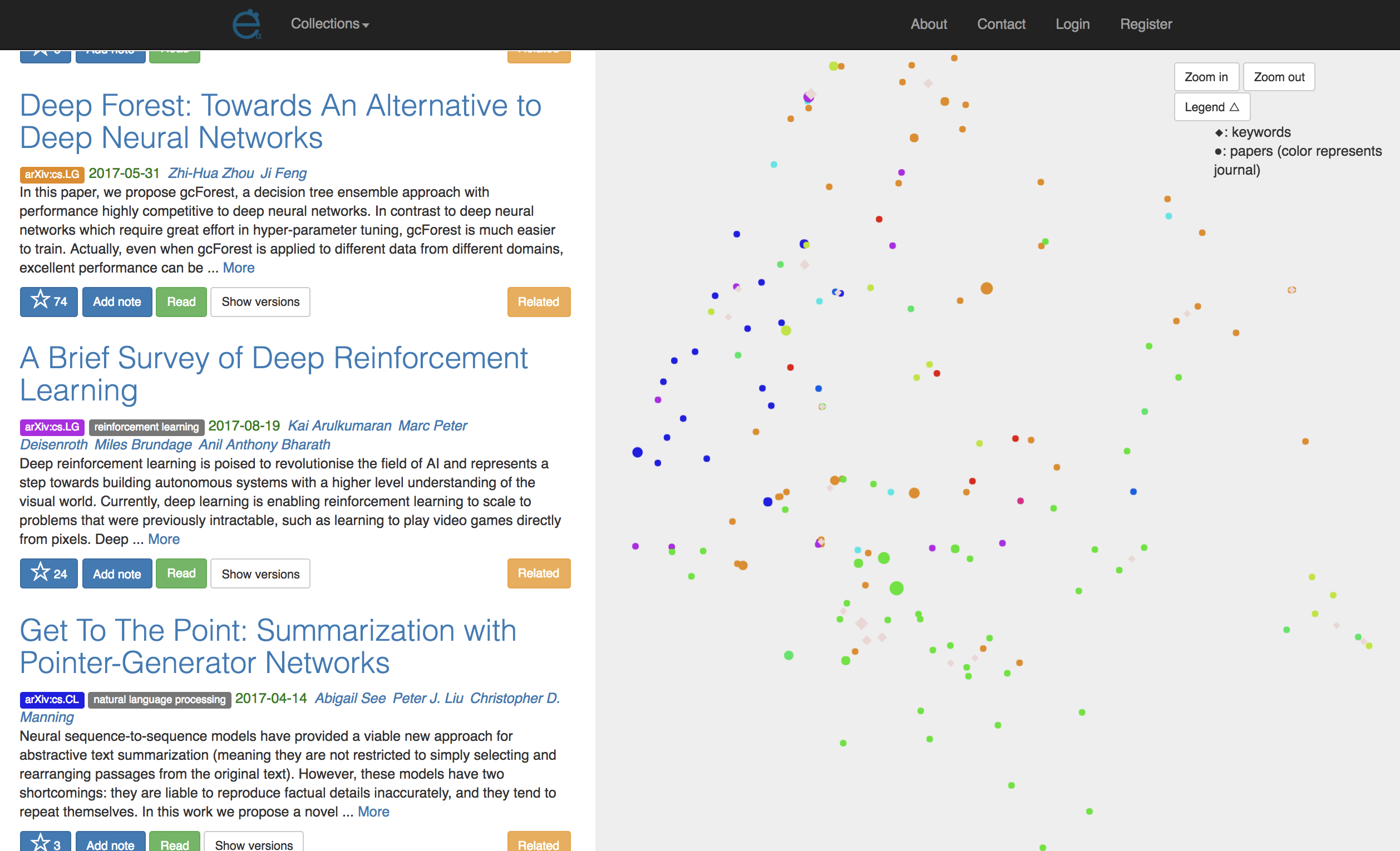}
\caption{The web interface of Etymo. The left part is a list of search results ordered by importance, while the right part is the corresponding visualisation of each paper in the list. Each paper is represented by a node and related papers are close to each other. The size of the node represents the importance and colour represents journal.}
\label{fig:etymo}
\end{figure}

A few recent scholar engines incorporate some form of visualisation in
displaying their search results.
AMiner
(\url{https://aminer.org/})\cite{tang08a} shows similar authors and the ego
network, which consists of a centre node (``ego'') and the nodes to whom the ego node is directly connected to.
AceMap~\cite{tan16} displays the
citation relationships between academic papers on a map,
in a similar way to Google Maps.
What we do differently is that we provide a combination of the traditional
list of results with a content similarity-based relationship visualisation.
Figure \ref{fig:etymo} shows the web interface of Etymo.

Why do we need a new interface? The most important reason is that it saves
our time in finding interesting research papers. The information we usually
need is the top ten papers from the search results and the papers related
to them, but there is no easy way to access all of that information at once
using the traditional list interface.  For example, to see related papers
of the current search result in Google Scholar, a user needs to click on
multiple 'related article' links.
  In Etymo, a user can check the top ten
rated papers on the list and locate all the related papers on the graph at
the same time.

%%%%%%%%%%%%%%%%%%%%%%%%%%%%%%%%%%%%%%%

\section{Experiments}

Etymo has over $36000$ papers in the database and we typically add $500$ new
papers every week.
The Analysis uses an instance of Amazon Elastic Compute Cloud (Amazon EC2)
m4.xlarge, which has $16$ vCPUs and $64$GB memory.
%Scalbility

We update our database on a daily basis. During each update, we need to
first find two sets of
% $1000$ dimensional
vector representations for all the newly
added papers using Doc2Vec and TF-IDF.
Training of both models are done on
a weekly basis.  We then use t-SNE to find the x,y location of all the
papers, i.e., reduce these $1000$ dimensional vectors to $2$ dimensional
vectors.
%% t-sne,
The computation of t-SNE can be done in $O(N\log N)$ and requires $O(N)$ memory \cite{van2014}, which makes it possible to learn embedding of data sets with millions of objects.
%% ranking, graph
The number of nodes in our similarity-based network is equal to the number of papers in our database. We apply PageRank and Reverse PageRank on this network.
The predominant method for computing the PageRank is the power method. At each iteration, we do a sparse matrix vector multiplication, which has complexity $O(pN)$, where $p$ is the average number of non-zero elements on every row of the matrix and $N$ is the dimension of the matrix. Usually $10$ iterations can produce a good approximate ranking \cite[Chap 8.2]{langville11}.

In general, we found that network-based ratings can improve search results
by highlighting historically important papers.  For the query ``t-sne'' (a popular machine learning algorithm for dimensionality reduction)
we show the top $5$ search results in the tables below. Table \ref{tab:reverse_pagerank} shows the results which
include the PageRank and Reverse PageRank ratings on the similarity-based
network, while Table \ref{tab:lexicon} does not.  We noticed that the one
with network-based ratings gives more weight to important papers. Note that
Maaten and Hinton's "Visualizing Data using t-SNE" is the original t-SNE
paper. Comparing with Google Scholar's search results in Table \ref{tab:google},
Etymo's top search results include more recent publications.

\begin{table}
  \caption {Top $5$ search results of the search query "t-sne". The results include a combination of PageRank and Reverse PageRank ratings.} \label{tab:reverse_pagerank}
\centering
    \begin{tabular}{p{3.3cm}|c|p{3.8cm}}
      Authors & Year & Title \\ \hline
      Laurens van der Maaten, Geoffrey Hinton & 2008 & Visualizing Data using t-SNE \\
      Laurens van der Maaten & 2014 & Accelerating t-SNE using Tree-Based Algorithms \\
      Yanshuai Cao, Luyu Wang & 2017 & Automatic Selection of t-SNE Perplexity \\
      George C. Linderman, Manas Rachh, Jeremy G. Hoskins, Stefan Steinerberger, Yuval Kluger & 2017 & Efficient Algorithms for t-distributed Stochastic Neighborhood Embedding \\
      Yukun Chen, Jianbo Ye, Jia Li & 2017 &
      Aggregated Wasserstein Metric and State Registration for Hidden Markov Models
    \end{tabular}
\end{table}

\begin{table}
\caption {Top $5$ search results of the search query "t-sne". The results
  do not include any network-based ratings.} \label{tab:lexicon}
\centering
\begin{tabular}{p{3.3cm}|c|p{3.8cm}}
  Authors & Year & Title \\ \hline
  Yanshuai Cao, Luyu Wang & 2017 & Automatic Selection of t-SNE Perplexity \\
  Laurens van der Maaten & 2014 & Accelerating t-SNE using Tree-Based Algorithms \\
  Maaten, Laurens van der, Hinton, Geoffrey & 2008 & Visualizing Data using t-SNE \\
  Richard R. Yang, Mike Borowczak & 2017 & Assessing Retail Employee Risk Through Unsupervised Learning Techniques \\
  Martin Renqiang Min, Hongyu Guo, Dinghan Shen & 2017 & Parametric t-Distributed Stochastic Exemplar-centered Embedding
\end{tabular}
\end{table}

\begin{table}
\caption {Top $5$ search results of the search query "t-sne" in Google Scholar.} \label{tab:google}
\centering
\begin{tabular}{p{3.3cm}|c|p{3.8cm}}
  Authors & Year & Title \\ \hline
  Maaten, Laurens van der, Hinton, Geoffrey & 2008 & Visualizing Data using t-SNE \\
  Laurens van der Maaten & 2014 & Accelerating t-SNE using Tree-Based Algorithms \\
  AR Jamieson ML Giger, K Drukker, H Li & 2010 & Exploring nonlinear feature space dimension reduction and data representation in breast CADx with Laplacian eigenmaps and t-SNE\\
  K Bunte, S Hasse, M Biehl, T Villmann & 2012 & Stochastic neighbor embedding (SNE) for dimension reduction and visualization using arbitrary divergences\\
  Maaten, Laurens van der, Hinton, Geoffrey & 2008 & [PDF] Visualizing Data using t-SNE \\
\end{tabular}
\end{table}

\section{Conclusion}

It is hard to quantify
new things. In research, the value of a newly
published work is usually unknown until citations become available.
The value of this work is to provide a new approach to improve search
results on new papers by exploiting the paper full text content and social
media data.
Our user interface combines the item list with item relationship visualisation, which saves researchers time in finding interesting research papers.

\bibliographystyle{ACM-Reference-Format}
\bibliography{strings,sigproc}

\end{document}